\documentclass[aps,prb,twocolumn]{revtex4}
\usepackage{graphicx}
\usepackage{epsfig}
\usepackage[dvips]{color}
\usepackage[english]{babel}
\usepackage[latin1]{inputenc}

\usepackage{times,mathptmx}

\usepackage{amsbsy,amsmath,amssymb,amsthm,amsfonts}
\usepackage{times,mathptmx}
\usepackage{bm}
\usepackage{textcomp}
\usepackage{units}
\usepackage{epsfig}
\newcommand{\comment}[1]{}
\newcommand{\eref}{Eq.\eqref}

\begin{document}

\date{\today}
\title{Excitonic Rayleigh scattering spectra of metallic single-walled carbon
nanotubes}

\author{Ermin Malic$^{1}$}\email{ermin.malic@tu-berlin.de}
\author{Janina Maultzsch$^2$}
\author{Stephanie Reich$^3$}
\author{Andreas Knorr$^1$}

\affiliation{
$^1$ Institut f{\"u}r Theoretische Physik, Nichtlineare Optik und Quantenelektronik von Halbleitern, Technische
Universit{\"a}t Berlin, 10623 Berlin, Germany\\
$^2$ Institut f{\"u}r Festk{\"o}rperphysik, Technische
Universit{\"a}t Berlin, 10623 Berlin, Germany\\
$^3$ Fachbereich Physik, Freie Universit{\"a}t  Berlin, 14195 Berlin, Germany}

\begin{abstract}
We have performed microscopic calculations of the Rayleigh scattering cross
section for arbitrary metallic single-walled carbon nanotubes. 
The focus of our investigations lies on excitonic effects and their influence on the characteristic features in a Rayleigh scattering spectrum.
Our approach is
based on
density matrix theory including tight-binding energies, the carrier-light
coupling as well as the carrier-carrier interaction. Due to the refractive index contribution to the scattering cross section, we observe
characteristic features  in Rayleigh spectra, such as a strong deviation from the Lorentz peak shape  and
 the larger oscillator
strength of the lower-lying transition $M_{ii}^-$ in the double-peaked structure,  independently of the chiral angle and the diameter of the investigated 
nanotubes.
We observe excitonic binding energies in the range of $\unit[60-80]{meV}$ for metallic nanotubes with diameters of $\unit[1.5-2.5]{nm}$.
The overlap of the excitonic transition with the close-by continuum has a significant influence on the peak shape and a minor influence on the peak intensity
ratios.
The presented
results are in good agreement with recent experimental data.

\end{abstract}

\maketitle
\section{Introduction}
Absorption, photoluminescence, and Raman scattering are standard spectroscopy methods 
to reveal optical properties of  nanoscale objects.\cite{reichbuch, jorio08} In particular, they 
have been applied to characterize carbon nanotubes (CNTs) of different chiral angle, diameter, 
and family.\cite{bachilo02,miyauchi04,telg04} In 2004, Sfeir at al.\cite{sfeir04}  introduced  
Rayleigh scattering as an additional powerful technique for identifying the optical finger print 
of individual single-walled CNTs.
It allows the
investigation of optical properties of \textit{individual} CNTs, since the weak
scattering signal is much easier to measure than e.g. the change in intensity
due to the absorption.  In combination with electron diffraction
data, Rayleigh scattering spectroscopy has successfully been applied to
determine the electronic structure of individual CNTs, in particular the
predicted peak splitting in metallic tubes due to the trigonal warping
effect\cite{saito00, reich00c} was proven experimentally.\cite{sfeir06}

Recently, excitonic effects in metallic
nanotubes have been experimentally proved by measuring their Rayleigh spectra.\cite{heinz07, berciaud10} 
Despite the large screening, metallic nanotubes were shown to exhibit binding energies around $\unit[50]{meV}$, which is small comparing to
semiconducting nanotubes,\cite{wang05b, maultzsch05c} but still larger than the thermal energy at room temperature. 
The experimental data on excitonic Rayleigh scattering spectra has not yet been complemented by
theoretical studies. In Refs.
\onlinecite{malic07b, malic08} we studied free-particle Rayleigh scattering
spectra of metallic and semiconducting CNTs showing characteristic features
in Rayleigh scattering spectra, which distinguish them from corresponding absorption spectra.
In this work, we address the question on how the formation of Coulomb-bound electron-hole pairs, i.e. excitons, influences these features. We perform
 investigations 
on (i) the excitonic transition and excitonic binding energy,  (ii) the trigonal warping splitting as a function of the diameter and the chiral angle, (iii) the relative oscillator
strength in the double-peaked spectra of metallic nanotubes, and (iv) the peak shape in Rayleigh spectra of metallic and semiconducting nanotubes.
Finally, we compare our results with recent experimental data.\cite{wu07, berciaud10}

\section{Rayleigh scattering cross section}
%First we give a derivation sketch\cite{bohren, sfeir04} for the Rayleigh scattering cross section $\sigma(\omega)$.
In our calculations, the Rayleigh scattering cross section is  considered for incident light
polarized along the nanotube axis accounting for the depolarization effect that
strongly suppresses light polarized perpendicular to the nanotube
axis.\cite{ajiki93}
Here, we briefly summarize the derivation of the Rayleigh
scattering cross
section:
Nanotubes are regarded as long cylinders with diameters small compared to the wavelength of light.
Starting from  Maxwell equations, the expression for $\sigma(\omega)$
can be derived by solving the scalar wave equation in cylindrical
coordinates,\cite{bohren}  and exploiting the limit of small nanotube radii.
The scattering cross section  is given by the ratio of the rate
\begin{equation}
W_s=\int_A\bm{S}_s\cdot \bm{e}_r \,dA\,,
\end{equation}
at which energy passes through the scattering surface $A$
and the incident irradiance. The rate $W_s$ is determined by the radial
component of the Poynting
vector of the scattered field  $\bm{S}_s=\frac{1}{2}\mathrm{Re}[\bm{E}_s \times
\bm{H}^*_s]$.
%The electric  and magnetic field satisfy the vector wave equation $\square \bm{E}(\bm r, t)=0$. Their components do not separately fulfill the scalar wave equation $\square \psi=0$.
 By introducing vector cylindrical harmonics\cite{bohren}
$
 \bm{M}=\nabla \times (\bm{e}_z\psi)$ and $\bm{N}=k^{-1}\nabla \times \bm{M}
$
 with a scalar function $\psi$, the wave number $k$, and the unit vector $\bm e_z$ parallel to the cylinder axis the problem can be simplified, since these functions satisfy both the vectorial and the scalar wave equation.
Once they are calculated, the electric and magnetic field $\bm E_s$ and $\bm H_s$  can be expanded in $\bm M$ and $\bm N$.
%with $\bm{E}=\sum^{\infty}_{n=-\infty}\left(A_n\bm{M}_n+B_n\bm{N}_n\right)$.
The scalar function $\psi$ is called a generating function for the vector
harmonics $\bm M$ and $\bm N$. Its choice depends on the investigated system,
its symmetries and boundary conditions.  For Rayleigh scattering from a
cylinder, the scalar function has to satisfy the wave equation in
cylindrical
polar coordinates.
An ansatz for the  solution  is
$
\psi(\rho,\phi,z)=Z_n(r)e^{in\phi}e^{ihz}
$
with $Z_n(r)$ as Bessel functions of first and second kind of integral order $n$ and with $r=\sqrt{k^2-h^2}$. The quantum number $h$ satisfies the boundary conditions between the cylinder and the surrounding medium.
Within the  limit of small particles, i.e. for cylinders with a diameter much
smaller than the wavelength of light ($\bm{k}\cdot\bm{r}\ll 1$ with $\bm k$ as
the wave vector of light and $\bm r$ as the radius of the cylinder),  the
scattering cross section $\sigma(\omega)$ per unit length can be expressed via
the linear susceptibility $\chi(\omega)$ as
\begin{equation}
\label{sigma}
\sigma(\omega) =\dfrac{\pi^2}{4c^3}r^4 \omega^3
|\chi(\omega)|^2\,,
\end{equation}
where  $\omega$ is the angular frequency of the light and $c$ the
speed of light. In contrast to the scattering from a sphere,
the cross section is proportional to $\omega^3$ instead of
$\omega^4$. This can be traced back to the scattered field $E_s$,
which is given asymptotically by $E_s \propto \frac{1}{\sqrt{k\,r}}$
far away from a cylinder, while for a sphere it is $E_s \propto
\frac{1}{k\,r}$.  The strength of the Rayleigh scattering is
determined by the square of the absolute value of the optical
susceptibility $\chi(\omega)$. In contrast to the absorption
coefficient\cite{malic06b} $\alpha\propto \mathrm{Im} \chi(\omega)$,
Rayleigh scattering has also a contribution from the real part of
$\chi(\omega)$ and, hence, includes the influence of the resonant
refractive index $n(\omega)\propto \mathrm{Re}\chi(\omega)$ of
optical transitions. This leads to important differences in the
characteristics of Rayleigh and absorption spectra, which are
discussed below.

\section{Density matrix approach}
To obtain the Rayleigh scattering cross section, we need the optical
susceptibility $\chi(\omega)$, which is
the linear response function  of the perturbed
system. Within the $\bm p \cdot \bm A$ approach for the light-matter
interaction, it reads\cite{scully}
\begin{equation}
\label{chi2} \chi(\omega) = \dfrac{j(\omega)}{\varepsilon_0
\omega^2 A(\omega)}
\end{equation}
with the externally driven current density $j(\omega)$ and the
vector potential $A(\omega)$.
The current density depends on the Fourier transform of the
microscopic polarization $p_{\bm k}(t)$ and the optical matrix element
$M_{vc}^z(\bm{k})$ along the nanotube axis (here, z-axis)\cite{malic06b,
hirtschulz08}
\begin{equation}
\label{j}
j(t)
=-i\dfrac{2e_0\hbar}{m_0}\sum_{\bm{k}}\mathrm{Re}\left(M_{vc}^z(\bm{k})p_{\bm{k}}
(t)\right).
\end{equation}
The microscopic polarization $p_{\bm k}(t)=\langle a^+_{\lambda \bm
k}a^{\phantom{+}}_{ \lambda' \bm k}\rangle (t)$  is a measure for the transition
probability between the two states $|\lambda \bm k\rangle$ and $|\lambda' \bm
k\rangle$, where $\lambda, \lambda'$ stand for the band index and $\bm k$ for
the
wave vector.  Our approach is formulated within the formalism of second
quantization with $a_{\lambda \bm k}$ and $a^+_{\lambda \bm k}$ as annihilation
and  creation operators.\cite{hirtschulz08}
As a result, the knowledge of $p_{\bm k}(t)$  allows the calculation of the
current density $j(\omega)$, which is required to obtain the optical
susceptibility $\chi(\omega)$. Finally, $\chi(\omega)$ determines the
Rayleigh scattering cross section $\sigma(\omega)$, cp. \eref{sigma}.

The temporal dynamics of $p_{\bm k}(t)$ is determined within the Heisenberg
equation of motion $
i\hbar \frac{d}{dt} p_{\bm k}(t)=[p_{\bm k}(t), H]
$
with the Hamilton operator
\begin{equation}
\label{H}
H=H_\text{0,c}+H_\text{c-f}+H_\text{c-c}\,,
\end{equation}
which   determines the dynamics of a physical system. 
The first two terms
describe the non-interacting carrier system in the presence of the external
electromagnetic field. In this work, a semiclassical approach is applied, i.e.
the charge carriers are treated quantum mechanically, while the field is
considered to be classical.
The carrier-field interaction reads
$
H_{\text{c-f}}=i\frac{e_0\hbar}{m_0}\sum_{\bm{l,l'}} \bm{M}_{\bm{l,l'}}\cdot \bm{A(t)}\, a^+_{\bm{l}}a^{\phantom{+}}_{\bm{l'}}
$
 with the optical matrix elements 
$
 \bm{M}_{\bm{l,l'}}
$, the vector potential $A(t)$, the electron mass $m_0$, and the elementary charge $e_0$.
The carrier-carrier interaction  is given by
$
 H_{\text{c-c}}=\frac{1}{2}\sum_{\bm{l}_1,\bm{l}_2,\bm{l}_3,\bm{l}_4} W^{\bm{l}_1,\bm{l}_2}_{\bm{l}_3,\bm{l}_4}a^+_{\bm{l}_1}a^+_{\bm{l}_2}a^{\phantom{+}}_{\bm{l}_4}a^{\phantom{+}}_{\bm{l}_3}
$
with the Coulomb matrix elements 
$ W^{\bm{l}_1,\bm{l}_2}_{\bm{l}_3,\bm{l}_4}$.
The single-particle energy $\varepsilon_{\bm{k}}$ required in the free carrier
contribution $H_{0,c}=\sum_{\bm{\bm k}} \varepsilon_{\bm{\bm k}} a^+_{\bm{\bm
k}}a^{\phantom{+}}_{\bm{\bm k}}$ is determined within  the
nearest-neighbor tight-binding (TB) approach.\cite{reich02b}

\begin{figure}[t!]
\center{\includegraphics[width=\columnwidth,angle
=0,keepaspectratio=true]{energy.eps}
}
\caption{Band
structure of the $(22,13)$ nanotube calculated with a) helical
quantum numbers $(\tilde{k}_z, \tilde{m})$  and b) linear quantum numbers
$(k_z,m)$. While there is only one helical subband $\tilde{m}$ with a large
Brillouin zone (BZ), 626 linear subbands $m$ with a small BZ need to be
calculated. The red dashed line shows the renormalized band structure due to the
electron-electron interaction. The arrows indicate the energetically
lowest optical
transitions, cp. Fig. \ref{spectra}.}\label{energy}
\end{figure}
\begin{figure}[b!]
\center{\includegraphics[width=0.75\columnwidth,angle
=0,keepaspectratio=true]{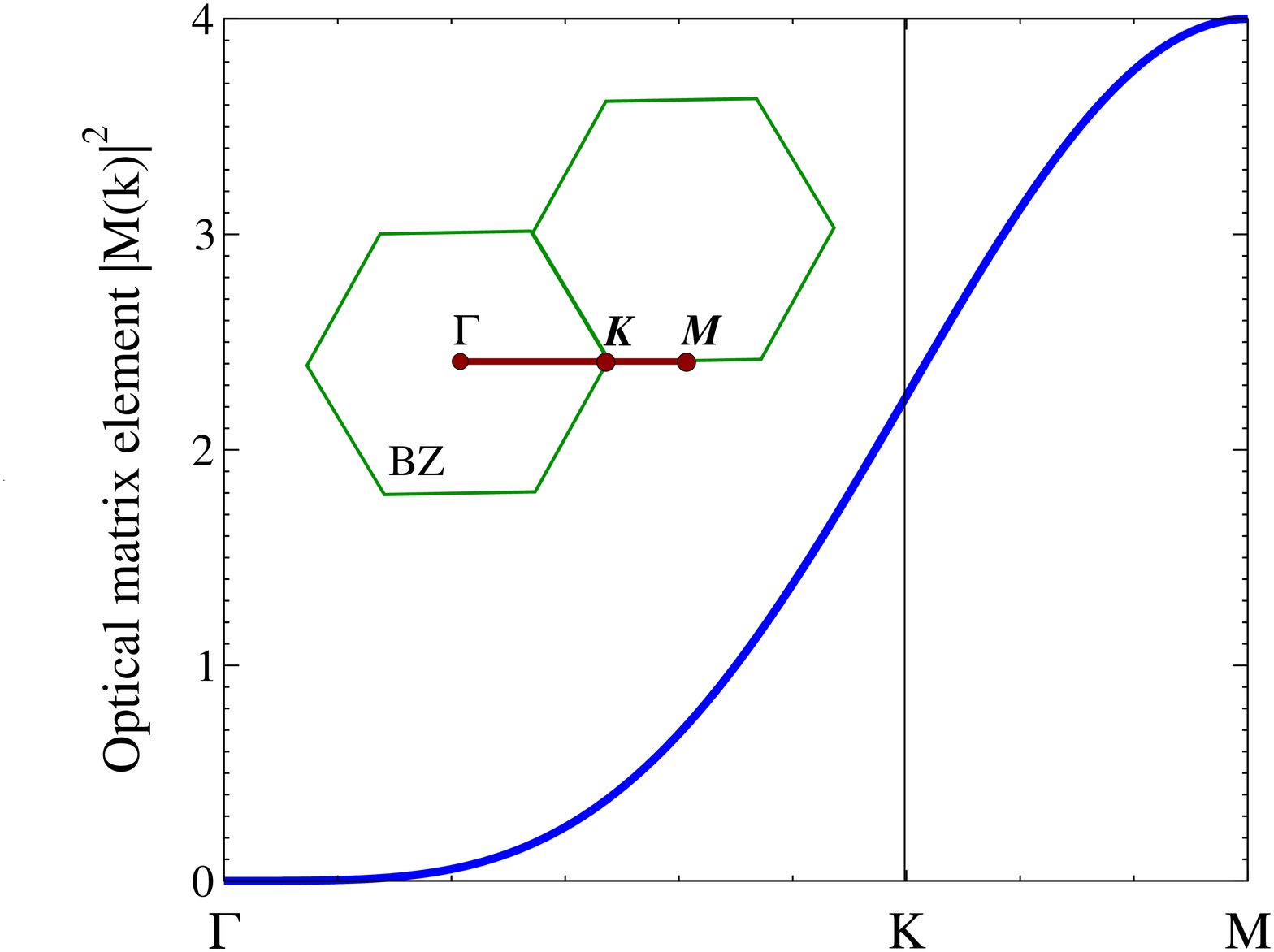}
}
\caption{Square of the absolute value of the optical matrix
element $M_{vc}(k)$ along the $\Gamma K M$ high-symmetry line in the Brillouin zone
of  graphene, cp. the inset.
Note, that the
matrix element has a smaller absolute value on the $k \Gamma$ than
on the $K M$-side. }\label{matrix}
\end{figure}

The periodic boundary conditions around the nanotube circumference are
considered by restricting the allowed wave vectors $\bm k$ to lines in the
graphene Brillouin zone (zone-folding
approximation).\cite{reichbuch}  The two-dimensional wave vector
$\bm{k}$ decouples in a continuous component $k_z$ along the nanotube axis and a
perpendicular quantized component $k_\perp=\frac{2}{d}m$ with the  diameter $d$
and the subband index $m$.
Since
nanotubes are described by line groups containing a screw axis,
two different sets of quantum numbers are possible: linear $(k_z, m)$ and helical
$(\tilde{k_z},\tilde{m})$ quantum numbers.\cite{damnjanovic03}  The linear $k_z\in(-\frac{\pi}{a},\frac{\pi}{a}\,]$ corresponds to the
pure translational subgroup of the line group. Here, $a$ stands for the translational period along the tube axis. The
linear momentum along the tube axis is a conserved
quantity. However,  the quasi-angular momentum $m\in(-\frac{q}{2},\frac{q}{2}]$ (with $q$ as the number of hexagons in the nanotube unit cells) contains
both pure rotations and screw axis operations. As a result, $m$ is
not fully conserved and Umklapp rules need to be taken into
account, when the Brillouin zone or the $\Gamma$ point are
crossed.\cite{reichbuch}
In contrast, the helical angular momentum $\tilde m\in(-\frac{n}{2},\frac{n}{2}\,]$ is a conserved quantity, since
it  corresponds to pure rotations of the nanotube. The number of 
helical subbands is considerably smaller compared to linear indices, see Fig.
\ref{energy}. The Brillouin zone, however,  is larger with
$\tilde{k}_z\in(-\frac{q}{n}\frac{\pi}{a},\frac{q}{n}\frac{\pi}{a}\,]$, where
$n$ is the greatest common divisor
of the chiral indices $n_1$ and $n_2$.  Figure \ref{energy} illustrates the two
different sets of quantum numbers by plotting the band structure of the metallic
$(22, 13)$ nanotube. In this work, we have applied helical indices taking all
subbands and the full Brillouin zone into account.

Note, that for nanotubes with small diameters hybridization effects might play an important role.\cite{blase94} 
Here, the zone-folded tight-binding wave functions can be inappropriate. In particular, these curvature effects have been shown to significantly
contribute to the wide family spread in Kataura plots.\cite{jiang07} Furthermore, the nearest-neighbor tight-binding approach is known
to be a good approximation for transitions close to the $K$ point, whereas it is often insufficient to model peak positions at higher energies. 
However, in our work
we focus on characteristic peak shapes and relative peak intensities in Rayleigh scattering spectra for metallic and semiconducting nanotubes, 
where we
expect hybridization effects to play a minor role.

The optical matrix element $M_{vc}(\bm k)=\langle \psi_v(\bm
k)|\nabla|\psi_c(\bm k)\rangle$, cp. Fig.\ref{matrix}, and the Coulomb matrix
element
$W^{12}_{~34}=\langle \psi_1 \psi_2|W_{Coul}|\psi_3\psi_4\rangle$ with the
screened Coulomb potential $W^{12}_{~34}$ enter into
the carrier-light Hamiltonian $H_{\text{c-f}}$ and the Coulomb Hamiltonian
$H_{\text{c-c}}$ in \eref{H}, respectively. They  are calculated analytically
by
applying the zone-folded tight-binding wave functions
$\psi(k)$.\cite{reichbuch}
Then, all necessary ingredients are available to determine the temporal evolution of the microscopic polarization $p_{\bm k}$ yielding\cite{hirtschulz08,malic08b}
\begin{eqnarray}
\label{bloch1}
\dot{p}_{\bm{k}}(t)=-i\tilde{\omega}_{\bm{k}} p_{\bm{k}}(t) +i \tilde{\Omega}_{\bm k}(t)-\gamma  p_{\bm{k}}(t).
\end{eqnarray}
 This Bloch equation is valid in the limiting case of linear optics, where
the driving field is considered to be small  resulting in negligible change in
occupation in valence and conduction band.\cite{haug} The dynamics of a system
is
fully determined by the microscopic polarization $p_{\bm k}$. The Coulomb interaction is considered within the Hartree-Fock level.\cite{hirtschulz08,malic08b}
The Rabi frequency
$$
 \tilde{\Omega}_{\bm k}(t)=\frac{e_0}{m_0}M^{cv}_z(\bm{k})A(t)-\frac{i}{\hbar}\sum_{\bm{k'}}W_{e-h}(\bm{k}, \bm{k'})  p_{\bm{k'}}
$$
  in \eref{bloch1} describes  the Coulomb renormalized strength of the electron-light interaction. The term includes the renormalization due to 
the attractive electron-hole interaction\cite{hirtschulz08,malic08b} $W_{e-h}(\bm{k}, \bm{k'})$. 
This term describes the formation of excitons. The strong Coulomb interaction in carbon nanotubes mixes the degenerate states at the $K$ and $K'$ point 
resulting in a partial lifting of the degeneracy and the formation of bright and dark excitonic states.\cite{jiang07} In the following, our investigations focus on the optically active (bright) states. 

The band gap energy
$$\tilde{\omega}_{\bm{k}}=\left(\omega_c(\bm{k})-\omega_v(\bm{k})\right)-\frac{i}{\hbar}\sum_{\bm{k'}}W_{e-e}(\bm{k},\bm{k'})
$$
 contains the renormalization due to the electron-electron coupling $W_{e-e}(\bm{k},\bm{k'})$ corresponding to the self-energy correction,
 see the red curve in Fig. \ref{energy}.
The Coulomb interaction is  screened within the static limit of the
Lindhard equation.\cite{haug} The Coulomb matrix elements are
calculated within the tight-binding approximation by introducing a
regularized Coulomb potential,  which is parametrized by the Ohno
potential.\cite{zhao04c,jiang07} More details can be found in Ref.
\onlinecite{malic08b}. A phenomenological parameter
$\unit[\gamma=(0.0125/\hbar)]{eV}$ is included into \eref{bloch1},
which determines the linewidth in the calculated
spectra.\cite{spataru04}  The parameter describes dephasing
processes resulting e.g. from electron-phonon interaction.
The influence of phonons and in particular the investigation of
phonon sidebands due to the exciton-phonon coupling is beyond the
scope of this work and will be in focus of future studies. However, there are simulations
on intersubband transitions predicting weak phonon satellites. \cite{butscher04}

The presented approach is similar to the Bethe Salpeter method\cite{jiang07} within the Hartree Fock level.
The advantage of the density matrix theory lies in particular in the description of the ultrafast relaxation dynamics of non-equilibrium
charge carriers allowing a microscopic access to their time and momentum-resolved scattering dynamics.\cite{malic09}

\section{Rayleigh scattering spectra}
\begin{figure}[t!]
\center{\includegraphics[width=\columnwidth,angle
=0,keepaspectratio=true]{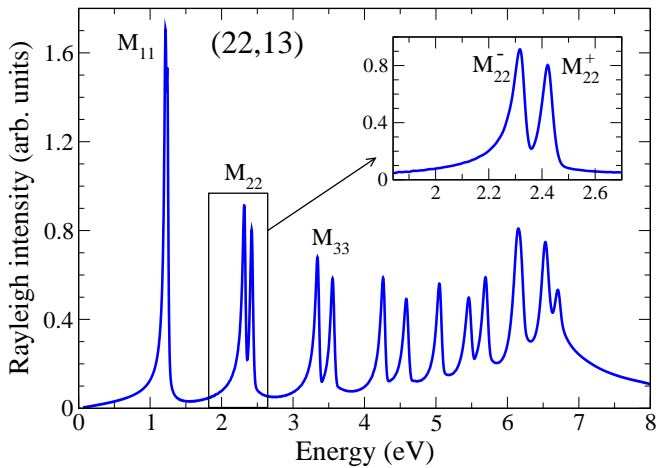}}
\caption{The figure shows the Rayleigh scattering spectrum for the exemplary
metallic $(22,13)$ nanotube. The inset is a blow-up of the third and forth
transitions $M^-_{22}$ and $M^+_{22}$ exhibiting characteristic features, such
as the peak
splitting and the asymmetry at the low-energy side of the transitions.
Note, that the first transition $M_{11}$ also shows a small
splitting, which can be resolved.}\label{spectra}
\end{figure}
Figure \ref{spectra} shows  exemplary the excitonic Rayleigh
scattering spectrum of the $(22,13)$ metallic nanotube. It is
characterized by a series of well-pronounced peaks stemming from
optical transitions between conduction and valence bands, cp.  the
arrows in Fig. \ref{energy}. For light polarized along the nanotube
axis, transitions are allowed between electronic states with $\Delta
m=0$ due to symmetry-imposed selection rules. As a result, the
absorption probability is particularly large for transitions between
the first valence band $v_1$ to the first conduction band $c_1$ at a
minimal energy $E_{11}$, followed by the transition $v_2 \rightarrow
c_2$ at $E_{22}$, etc. The corresponding peaks in the spectrum of
metallic tubes are denoted with $M_{11}, M_{22}$, etc.
Figure \ref{spectra} illustrates several characteristic
features of Rayleigh scattering spectra of metallic carbon
nanotubes: (i) a pronounced double-peaked structure of the optical
transitions due to the trigonal warping effect, (ii) stronger intensity
of the lower-lying transitions, i.e. the
oscillator strength of $M_{ii}^-$ is larger than of $M_{ii}^+$, and
(iii)  an asymmetry towards lower energies corresponding to an enhanced
cross section $\sigma(\omega)$ at the lower-energy
wing.

\begin{figure}[t!]
\center{\includegraphics[width=0.8\columnwidth,angle
=0,keepaspectratio=true]{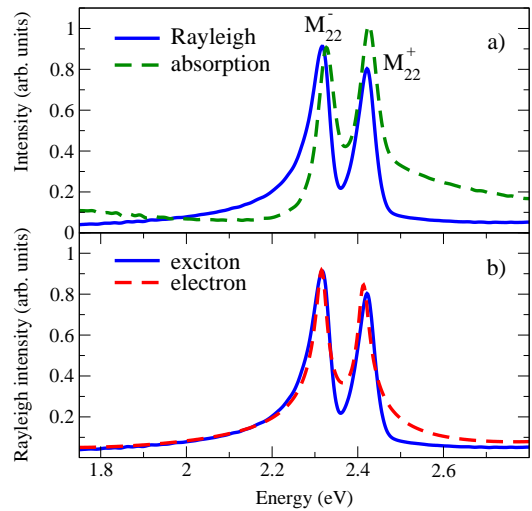}
}
\caption{a) Comparison between the peak shape in the excitonic Rayleigh scattering
and the excitonic absorption spectrum of the exemplary $(22,13)$ nanotube. The
absorption intensity is normalized to the $M^-_{22}$ Rayleigh peak. 
Note the different behavior of the intensity ratio in Rayleigh and absorption spectra. b)
Illustration of the peak shape within the free-particle  and the excitonic calculation of the Rayleigh
scattering cross section of the $(22,13)$ nanotube. The figure shows the
transitions $M^-_{22}$ and $M^+_{22}$. Since only the peak shape and the relative
intensity are in the focus of investigation, the free-particle spectrum is
rescaled in intensity and shifted in energy according to the
excitonic shift.}
\label{exc-el}
\end{figure}

Figure \ref{exc-el}a) shows a comparison between the
Rayleigh scattering spectrum and the absorption spectrum of the
exemplary metallic (22,13) nanotube. The largest difference is obtained with
respect to the peak shape. The absorption peaks are Lorentzians
reflecting the dependence of the absorption coefficient on
$\mathrm{Im} \chi(\omega)$. In contrast, the shape of Rayleigh
peaks is more complicated showing deviations from the Lorentzian
shape on both lower and higher energy side. This can be explained by
the interference with the real part of the optical susceptibility,
since the Rayleigh scattering cross section is given by
$|\chi(\omega)|^2$, as discussed below in detail. Furthermore, the peaks are slightly red-shifted
and the intensity ratio is reversed compared to the absorption
spectrum.

Figure \ref{exc-el}b) shows the difference between the
excitonic and the corresponding free-particle Rayleigh spectrum of
the $(22,13)$ nanotube.
 The figure illustrates the excitonic effects on Rayleigh scattering spectra:
(i) a considerable blue-shift of the free-particle transition
energies, as already shown for absorption spectra\cite{malic09,
hirtschulz08, malic08b}, (ii) the asymmetry towards lower energies
remains unchanged, when excitonic effects are included, (iii) the
intensity ratio of the double-peaked structure is slightly
increased, and (iv) the cross section at the higher-energy side of
transitions is reduced. 
In the following paragraphs, the observed characteristic
features of excitonic Rayleigh scattering spectra are discussed in
detail.

\subsection{Excitonic binding energies}

Excitonic effects significantly influence optical properties
of carbon nanotubes, as shown for absorption spectra in many
previous reports.\cite{spataru04, perebeinos04, mol05, capaz06,
jiang07, deslippe07, hirtschulz08, malic09} Excitonic binding
energies in the range of \unit[300-400]{meV} have been observed for
semiconducting\cite{wang05b, maultzsch05c} and in the range
of \unit[100]{meV} for metallic nanotubes.\cite{heinz07} Our
investigation on Rayleigh scattering spectra are in good agreement with these findings.
We observe strong shifts due to the formation of bound electron-hole
pairs. The binding energies are around
\unit[60-80]{meV} for investigated metallic nanotubes with $d \approx \unit[1.5-2.5]{nm}$. For semiconducting nanotubes, we observe 
excitonic binding energies in the range of \unit[200-400]{meV}.

Our approach allows the investigation of a large variety of different carbon nanotubes. The calculation of the Kataura plot reveals 
the diameter and the chirality dependence
of the excitonic transition and binding energy. As already reported in literature,\cite{jiang07} it
exhibits main $1/d$ lines for each transition $M_{ii}$ and a
characteristic V-shaped structure reflecting the chirality dependence of the trigonal-warping splitting.

\subsection{Trigonal warping peaks splitting} 
\begin{figure}[t!]
\center{\includegraphics[width=\columnwidth,angle
=0,keepaspectratio=true]{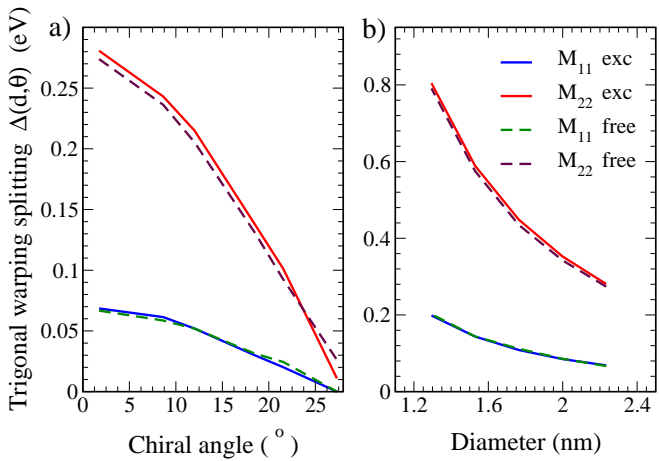} }
\caption{Trigonal warping splitting $\Delta(d,\theta)$ as a
function of a) the chiral angle $\theta$ (at a nearly constant diameter $d_0\approx \unit[2.3]{nm}$) and b) the diameter $d$ (at a nearly constant chiral angle
$\theta_0\approx 2-3^o$) for
metallic nanotubes.}\label{trig}
\end{figure}

Trigonal warping describes
 the deviation of the equi-energy contours from circles in
the Brillouin zone of graphene around the $K$
point.\cite{saito00,reich00c} 
 Due to its
trigonal shape, an energy splitting of  Van-Hove singularities stemming from different sides with respect to
the $K$ point appears. This strongly depends
on the orientation of the triangle:
The splitting is maximal for nanotubes with a small chiral angle and it vanishes for armchair tubes.
Furthermore, the higher the transition energy, the larger is the
trigonal warping effect, since the deviation from circles is larger. Figure \ref{trig} shows the peak splitting
$\Delta(d,\theta)$ as a function of the chiral angle and the
diameter for metallic nanotubes for both the excitonic and
the free-particle picture. First, we observe that excitons do not
influence the trigonal warping induced splitting. Second, we find that the
splitting  scales with diameter $d$ as $\Delta(d,\theta_0) \propto
A/d^2$ at a constant chiral angle $\theta_0$. The coefficient $A$
only depends on the order of the transition. The larger the diameter,
the smaller is the transition energy and the weaker is the trigonal
warping effect.
 For the dependence on
the chiral angle $\theta$, the scaling law is $\Delta(d_0,\theta)
\propto A_1-A_2 \theta^2$ at a nearly constant diameter $d_0$ confirming that the splitting is maximal for zigzag nanotubes with $\theta=0^o$
 and zero for armchair tubes  with $\theta=30^o$.

\subsection{Peak intensity ratio} 
 The lower-lying transitions within the double-peaked structure of Rayleigh scattering spectra
show a higher oscillator strength independently of the chiral angle
and diameter, i.e. $M_{ii}^-$ is always higher in intensity than
$M_{ii}^+$. The intensity ratio $R_{ii}=I(M_{ii}^-)/I(M_{ii}^+)$ increases
with decreasing chiral angle. In the limiting case of armchair
nanotubes, the ratio is exactly 1 due to vanishing splitting. The
described behavior of peak intensity ratios is not significantly influenced by
excitons.

 The relative intensities can be explained by the different
behavior of the optical matrix element $M_{vc}(k)$ entering in
$H_{\text{c-f}}$ along the two high-symmetry lines $K\varGamma$ and
$KM$ in the graphene BZ.\cite{malic08} The carrier-field interaction
turns out to be higher on the $KM$ side, cp. Fig. \ref{matrix}. As a
result, the lower-lying transitions in the double-peaked structure
stemming from the $KM$ side\cite{thomsen07} are amplified. Following
this argumentation, the intensity ratio should increase with the
order of transition. However, the dependence of the scattering cross
section $\sigma(\omega)$ on $\omega^3$ cancels this effect, since
the energetically higher transition $M_{ii}^+$ is enhanced with
respect to $M_{ii}^-$ resulting in a decrease of the intensity
ratio.

Another interesting observation is the inverse intensity
ratio $R_{ii}<1$ in absorption spectra, cp. Fig. \ref{exc-el}a). This can be ascribed to the overlap
of the $M_{ii}^+$ peak with the 
Van-Hove singularity associated with $M_{ii}^-$. The  high-energy tail
of the Van-Hove singularity enhances the intensity of $M_{ii}^+$ leading to an intensity ratio $R_{ii}$ smaller than 1. 
In the case of Rayleigh scattering, the overlap with the continuum is smaller, since here the continuum is not characterized by a Van Hove singularity.
 As a result,  the intensity ratio $R_{ii}$ remains larger
than 1 - as expected from the family behavior of the optical matrix element.

\begin{figure}[b!]
\center{\includegraphics[width=\columnwidth,angle
=0,keepaspectratio=true]{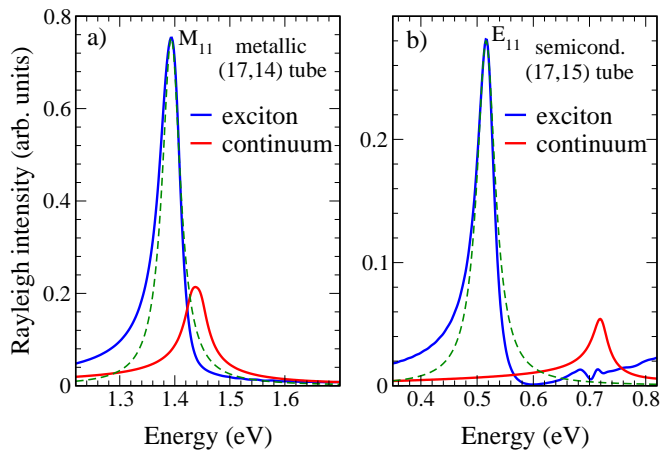} }
\caption{Rayleigh scattering spectra of an exemplary
a) metallic and b) semiconducting nanotube. The figure shows the excitonic transition and the continuum (containing the 
bandstructure renormalization due to the electron-electron coupling). The peak shape is fitted
with a Lorentzian with a width of approximately 40\,meV (dashed
lines). }\label{ray}
\end{figure}

\subsection{Peak
shape}
 
\begin{figure}[t!]
\center{\includegraphics[width=\columnwidth,angle
=0,keepaspectratio=true]{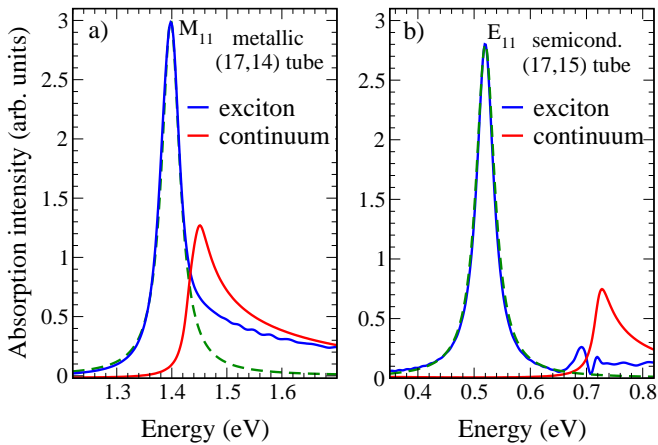} }
\caption{The same as in Fig. \ref{ray}, but showing excitonic and free-particle absorption spectra of an exemplary a) metallic and b)
semiconducting nanotube. 
The figure also exhibits the formation of higher excitonic states in the spectra of
semiconducting CNTs corresponding to the Rydberg series in the hydrogen atom.}\label{abs}
\end{figure}

Figures \ref{ray} and \ref{abs}  illustrate the characteristic 
peak shape of an exemplary
metallic and a semiconducting nanotube in excitonic (solid blue lines) and free-particle (solid red lines) Rayleigh scattering and absorption spectra, 
respectively.
For comparison, the figures also show a fit with a Lorentzian  in the background (dashed green lines). 
Rayleigh peaks are shown to be
asymmetric towards lower energies reflecting an enhanced cross
section $\sigma(\omega)$ at the lower-energy wing, cp. Fig. \ref{ray}. This can be
traced back to the refractive part of the optical susceptibility.
The latter exhibits a long tail on the low energy side of each
transition, which adds up with the resonant response leading to the
observed asymmetry. Both metallic and semiconducting nanotubes
exhibit this characteristic asymmetry leading to a considerable
broadening of the ''Lorentzian-like'' Rayleigh peaks\cite{berciaud10} - in contrast to the corresponding peaks in the absorption spectra, cp. Fig. \ref{abs}.

Furthermore, we observe an interesting feature on the
high-energy side of metallic nanotubes. Here, two effects are
competing: On the one side, the spectrally decaying refractive index leads to a
reduction of the scattering cross section. On the other hand, due to
the small binding energies there is an overlap between the excitonic
transition and the continuum lifting up the intensity. As a result,
the overall reduction is much smaller compared to semiconducting nanotubes,
where the excitonic binding energy is large and the overlap
with the continuum is negligibly small.

For comparison, Fig. \ref{abs}  shows the peak shape in excitonic and free-particle 
absorption spectra for the same exemplary metallic and semiconducting nanotube as in Fig. \ref{ray}. Since the
absorption coefficient is determined only by the imaginary part of
the optical susceptibility, the asymmetry to lower energies and the
resulting broadening are not present in absorption spectra. The
peaks are perfect Lorentzians (dashed line) reflecting the excitonic
character of the transition. There is no interference with the refractive part of the response function resulting in narrow peaks with a width 
of \unit[40]{meV} 
determined by the parameter $\gamma$ entering \eref{bloch1}.
Note that the peak shape of metallic nanotubes exhibits a higher-energy shoulder
due to the small excitonic binding energies and the resulting overlap with the continuum.\cite{deslippe07}

\begin{figure}[t!]
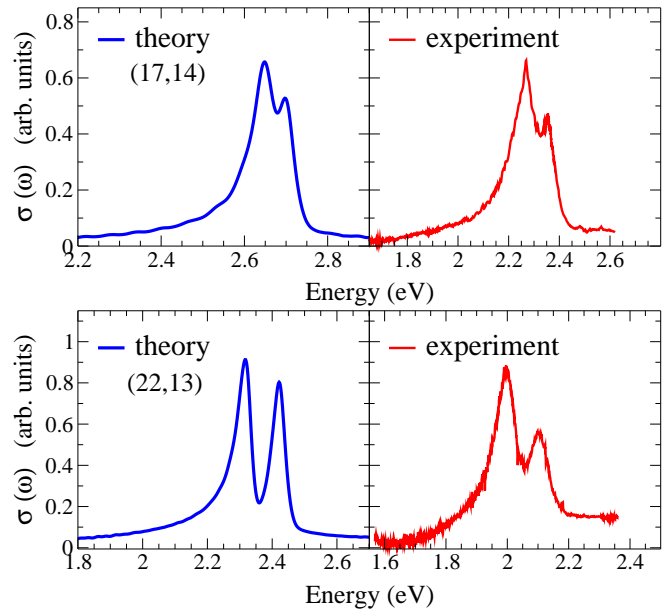

\center{\includegraphics[width=\columnwidth,angle
=0,keepaspectratio=true]{th_exp_17-14.eps}
\includegraphics[width=\columnwidth,angle
=0,keepaspectratio=true]{th_exp_22-13.eps}
}
\caption{Comparison between the experimental and theoretical
excitonic Rayleigh scattering spectra for the $(17,14)$ and the  $(22,13)$
metallic nanotubes.
The experimental results are taken from Ref. \onlinecite{wu07}.
 }\label{exp-th}
\end{figure}
\subsection{Comparison to experiment}
  Figure \ref{exp-th} illustrates the good
agreement of theoretical and experimental Rayleigh scattering spectra for two
exemplary metallic tubes. Including excitonic effects  even further improved the
comparison with the
experiment.\cite{malic08, wu07} As predicted in theory, the experimentally
observed Rayleigh spectra show a double-peaked structure with a clearly enhanced
scattering intensity at the  lower energy wing of transitions. Furthermore, the
oscillator strength of the first peak in the double-peaked structure is found to
be stronger in intensity. This agrees well with the experiment, where the
intensity ratio is even more pronounced.
The calculated transition energies, however, differ from the experimental
results. They are blue-shifted by approximately $\unit[0.3-0.4]{eV}$ compared to
the experiment. This deviation can be traced back to the calculation of the band
structure within the nearest-neighbor tight-binding approach,  which is
known to be a good description for transitions close to the $K$
point.\cite{reichbuch} For higher energies, however,  a considerable blue-shift
in comparison to third-nearest neighbor TB or first-principle calculations
was observed.\cite{reich02b}
Extensions to third-nearest neighbor TB or extended TB
calculations\cite{popov04b, jiang07} would further improve the comparison with
experimental data.

\section{Conclusions}
We have performed microscopic calculations of the Rayleigh scattering cross section including excitonic effects for arbitrary metallic
 single-walled carbon
nanotubes. Our approach is based on the density matrix formalism combined with zone-folded tight-binding wave functions. 
While the absorption coefficient is given only by the imaginary part of the optical susceptibility, the Rayleigh scattering cross section
also contains the influence of the real part corresponding to refractive index contribution. This leads to characteristic features  in Rayleigh
scattering spectra, such as the strong deviation from the Lorentz peak shape exhibiting an enhanced cross section on the lower energy wing, and
 the larger oscillator
strength of the lower-lying transition $M_{ii}^-$ in the double-peaked structure  independently of the chiral angle and the diameter of the investigated tubes.
We discuss the influence of excitonic effects on these characteristic features  including a study on the trigonal warping splitting.
The
comparison with recent experimental data yields a good agreement with respect
to the characteristic peak shape and the peak intensity ratios.\\

We acknowledge the support from Sfb 658 and the ERC under grant number
210642. Furthermore, we thank T. F. Heinz (Columbia
university) for fruitful discussions.

%\bibliographystyle{apsrev}
%\bibliography{/home/ermin/data/paper-conferences/papers.bib}

\end{document}